# DEVELOPMENT OF A PLASMA SIMULATION TOOL FOR ACCELERATING CAVITIES *

N. K. Raut†, I. H. Senevirathne, T. Ganey, P. Dhakal, and T. Powers
Thomas Jefferson National Laboratory, Newport News, VA 23606, USA

*Abstract*

Plasma processing of superconducting radio frequency (SRF) cavities has shown an improvement in accelerating gradient by reducing the radiation due to field emission and multipacting. Plasma processing is a common technique where the free oxygen produced by the plasma breaks down and removes hydrocarbons from surfaces. This increases the work function and reduces the secondary emission coefficient. The hydrocarbon fragments of $H_2$, $CO$, $CO_2$, and $H_2O$ are removed from the system with the process gas which is flowing through the system. Here, we present COMSOL for the first time to simulate the plasma processing of an SRF cavity. In this work, we use Jefferson Lab's C75 SRF cavities design as our case study. Using simulation, we predict the condition of plasma ignition inside the SRF cavity. The simulation provides information about the optimal rf coupling to the cavity, mode for plasma ignition, choice of gas concentration, power, and pressure.

## INTRODUCTION

Superconducting radio frequency (SRF) cavities made from niobium are the building blocks for many modern particle accelerators around the world. The SRF cavities store the electric fields that provides the required acceleration to charged particles. These high electric fields are concentrated in a small confined enclosure, thus increasing the probability of both electron emission from the cavity surface and their subsequent acceleration. These electrons gain energy and guided by the magnetic fields, impact the cavity surface producing electromagnetic radiation, which in turn, further degrades the performance of SRF cavities. Field emission and multipacting are commonly observed in SRF cavities if the cavity surface contains sharp features or the surface is contaminated [1]. Recent developments have shown that performance degradation can be minimized by plasma processing with a mixture of gas ignited by microwave power [2]. The plasma disintegrates most of the organic contaminates and is pumped out from the cavity enclosure [3].

A mixture of inert gas (Ar, He, Ne) with a concentration of >80% and a lower concentration (∼1-20%) of highly reactive oxygen is used to create plasma inside the cavity by an exchange of energy between the electromagnetic field of the cavity and the gas mixture. The free oxygen species created during this process then react with a hydrocarbon as $O_2 + C_xH_y \rightarrow CO_2 + CO + H_2O$ breaking down the organic bonds. This process increases the work function ($\phi$) and decreases the secondary emission yield (<SEY>) of the surface of the metal [4]. The $Ar/O_2$ gas ionizes and creates plasma when the gain of energy by the gas mixture from the cavity's field is >15.8 eV, which produces highly reactive free oxygen species. The $O_2$ plasma then oxidizes hydrocarbons from the contaminated cavity's surface. Such volatile by-products are continuously pumped out from the system.

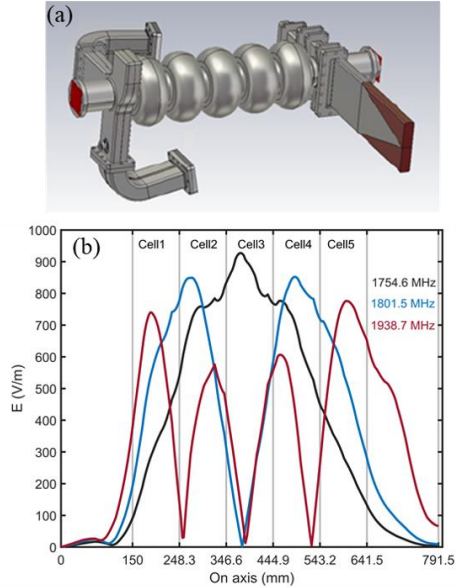

Figure 1: (a) Configuration of the C75 cavity used in this simulation, (b) the electric field profile of three dipole modes: 1754.6 MHz, 1801.5 MHz, and 1938.7 MHz used in the simulation.

Plasma processing has been applied to different shapes of elliptical cavities resonating at 0.8 GHz at Oak Ridge National Lab [5, 6], 1.3 GHz cavities at Fermi Lab [7], and 1.5 GHz cavities at Jefferson Lab [8]. In all these cases the rf power is injected through the higher-order mode coupler. The similarities between the higher order mode couplers installed in these cavities made it easier to couple the cavity with microwave power.

Significant progress has been made in the plasma processing of SRF cavities to improve the performance, as well as saving the costs associated to refurbishment of under-performing cryomodules. A recent report from Jefferson Lab on the plasma processing of four C100 cryomodules has resulted in 49 MeV operational energy gain. An effort has been made to process other types of accelerating cavities installed in the CEBAF accelerator. One of the new designs, namely C75 employs a waveguide higher order mode with load installed inside the enclosure, which complicates the process of plasma processing that was developed earlier. In this manuscript, we used COMSOL Multiphysics simulation to ignite the plasma inside the C75 cavities, in particular, to identify the:

- Optimum $S_{11}$ for the plasma ignition.
- Choice of $O_2$ concentration, pressure, and power.

* This work is supported by SC Nuclear Physics Program through DOE SC Lab funding announcement Lab-20-2310 and by the U.S. Department of Energy, Office of Science, Office of Nuclear Physics under contract DE-AC05-06OR23177.
† raut@jlab.org

## MICROWAVE SIMULATION

As a cost-saving measure, the C75 program reused the end groups off of a C20 cavity that was removed from the CEBAF accelerator [9]. The cavity end groups consist of a beam tube and rectangular fundamental power coupler waveguide on one end and two rectangular high order mode (HOM) damping waveguides and stubs on the order side, as shown in Fig. 1(a). Due to this coupling design, the developed plasma processing technique needed to be revised.

Microwave simulations were done to identify the dipole modes with suitable electric field profiles within the C75 cavity. As shown in Fig. 1(b), three dipole modes resonating at 1754.6 MHz, 1801.5 MHz, and 1938.7 MHz were identified from simulations. When the cavity resonates at 1754.6 MHz, the field profile has the highest electric field on cell 3, whereas the peak electric field shifts towards cells 2 and 4 at a frequency of 1801.5 MHz. At the next higher frequency, 1938.7 MHz, the peaks in the electric field shift towards the end cells.

The scattering (S) parameter is a key parameter that allows us to identify the amount of rf power coupled to the cavity. For the plasma ignition, it is important to have high energy within the cavity compared to the reflected and transmitted energy. The S11 of the cavity is related to the quality factor ($Q_0$) as:

$$Q_0 = \frac{2P_F}{P_{diss}} \left(1 + C_\beta |S11|\right) Q_L \quad (1)$$

Here, $P_F$ is the forward power, $Q_L$ is the loaded quality factor and, $C_\beta$ is +1 if the cavity is over coupled and -1 if it is under coupled. The value of $Q_0$ quantifies the ability of the cavity to store energy compared to losses in one RF cycle given by a relation:

$$Q_0 = \frac{\omega U}{P_{diss}} \quad (2)$$

where ω is the angular frequency, U is the stored energy, and $P_{diss}$ is the dissipated power into the walls of the cavity. S11 directly affects the energy-storing ability of the cavity. In the state of the critical coupling, the cavity can store almost all energy sent into the cavity. The electric field in the cavity scales as the square root of the stored energy [4]. Higher electric fields produce plasma with an increased electron density during plasma processing.

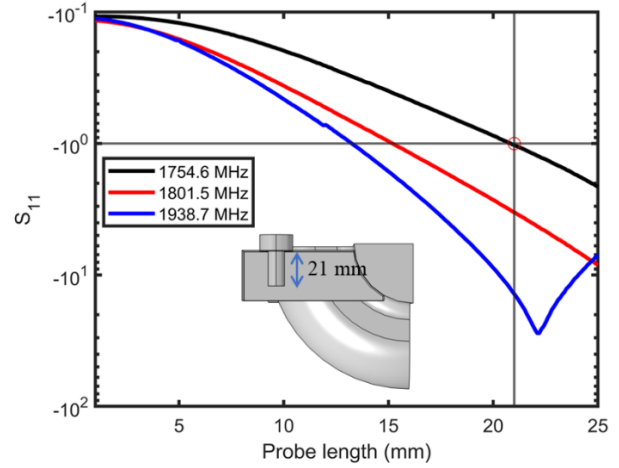

Figure 2: S11 of the cavity as a function of the probe length for three modes (a) 1754.6 MHz, (b) 1801.5 MHz, and (c) 1938.7 MHz of the cavity.

## MULTIPHYSICS SIMULATION

We used COMSOL® simulation software for the plasma simulation [10]. The transport properties of electrons are calculated using drift-diffusion equations [11]. Frequency domain simulations were used to excite the cavity modes. For the simulations, we used mixtures of argon-oxygen gas. A total of 40 reactions (5 of Ar and 35 of $O_2$) were used. The reactions were elastic, attachment, excitation, and ionization in nature. They were exported from the open-source database system LxCat [12].

## RESULTS AND DISCUSSION

### Choice of Coupling Probe

In the initial simulations, the power into the cavity was adjusted by means of changing the RF coupling strength. However, with the cavity as assembled in cryomodules, the coupling strength was optimized for minimum RF power at the fundamental frequency based on operating gradient, beam loading and microphonics [9]. Further, the input coupler was designed to couple out HOM power generated in the TE111 modes, which are below the cutoff frequency of the HOM waveguide. The magnitude of S11 at a given frequency is a metric of the amount of power that goes into the cavity. In the C75 configuration, a copper probe of diameter ~5 mm was inserted in the input top hat waveguide installed on the fundamental power coupler (FPC). Figure 2 shows S11 of three modes (1754.6 MHz, 1801.5 MHz, and 1938.7 MHz) as a function of the probe length. The length of the probe is referenced from the surface to the volume of the waveguide as shown in the inset of Fig. 2. For a probe length of ∼21 mm S11 was found to be -1 dB, -3 dB and -9 dB for modes 1754.6, 1801.5, and 1938.7 MHz respectively. Although the experimental setup may vary, we have fixed the probe length of 21 mm for plasma simulation.

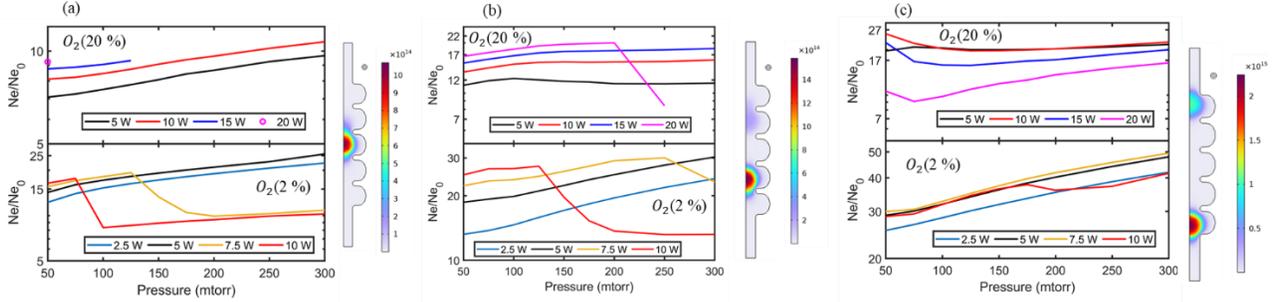

Figure 3: Fractional change in the electron number density at 2% and 20% of oxygen concentration at different pressures and power for (a) 1754.6 MHz, (b) 1801.5 MHz, and (c) 1938.7 MHz. The initial number density ($Ne_0$) is set at $10^{14}$.

*Oxygen Concentration and Microwave Power*

A key parameter in the plasma simulation is the change in the electron number density (Ne). During the plasma ignition process. The background electrons are accelerated by the RF fields until they are able to ionize the gas molecules. A cascade process occurs and the plasma is ignited. Figure 3 shows a fractional change in Ne from the initial number density ($Ne_0$) set at $10^{14}$, at different pressures and powers, for 2% and 20% oxygen concentration. For both oxygen concentrations, we ran simulations from 50-300 mtorr pressure. Higher power (5-20 watts) was required to ignite the plasma for 20% oxygen, whereas lower power (2.5-10 watts) was required to ignite a 2% oxygen/argon gas mixture [13].

For the 20% oxygen mixture, we observed growth in the number density (Ne) as a function of the pressure at 5 W and 10 W for all three modes. However, as we increased the power, the plasma migrated away from the source of RF and towards the FPC. This occurred when more than 15 W was applied to the system at 1754.6 MHz. Plasma was produced in two cells simultaneously when two higher order mode frequencies were applied. This is consistent with experimental observations using a camera on either end of the cavity [14].

For 2% oxygen concentration, simulations were done at input powers of 2.5, 5, 7.5, and 10 W. The electron number density at powers 2.5 W and 5 W at 2% oxygen is higher than that at any power and pressure at 20% oxygen. Here, the lower power <7.5W is good enough for plasma ignition [15].

## SUMMARY

We have developed a simulation tool to optimize rf coupling, choice of operating modes, choice of gas concentration, power, and pressure for plasma processing accelerating cavities. One of our findings is for Ar/$O_2$ plasma, 2% oxygen concentration could be as effective as 20% oxygen concentration at the lower powers. We plan to benchmark simulation results with experiments in the C75 cavities at Jefferson Lab.